\documentclass[10pt]{ietbook}

\usepackage{xcolor}
\usepackage[official]{eurosym} 
\usepackage{ulem}
\usepackage{etoolbox} % For if statement (toggle comments) 
\renewcommand \thesection {\arabic{section}}
\renewcommand\thesubsection   {\thesection.\arabic{subsection}}
\renewcommand\thesubsubsection{\thesubsection .\arabic{subsubsection}}

\begin{document}

%For RH Book title
\rhbooktitle{Privacy by Design for the Internet of Things}

\markboth{Privacy by Design for the Internet of Things}{Towards an accountable Internet of Things: A call for reviewability}

%\headname{\begin{footnotesize}\color{red}Published in \emph{IEEE Access}, vol. 7, pp. 6562--6574, 2019. DOI: %\href{https://doi.org/10.1109/ACCESS.2018.2887201}{10.1109/ACCESS.2018.2887201}\end{footnotesize}}

\cauthor{Chris Norval,
Jennifer Cobbe, %and %\thanks{Author affiliations} and\\
Jatinder Singh \thanks{Compliant and Accountable Systems Group, Department of Computer Science \& Technology, University of Cambridge, UK. Contact: firstname.lastname@cst.cam.ac.uk}} %\thanks{Author affiliations}}

\newcommand{\notecn}[1]{{\color{teal}[{\bf Chris:} #1]}}
\newcommand{\changecn}[1]{{\color{teal}#1}}	
\newcommand{\notejs}[1]{{\color{purple}[{\bf Jat:} #1]}}
\newcommand{\changejs}[1]{{\color{brown}#1}}
\newcommand{\notejc}[1]{{\color{blue}[{\bf Jen:} #1]}}
\newcommand{\changejc}[1]{{\color{blue}#1}}
\newcommand{\todo}[1]{{\color{red}[{\bf To Do:} #1]}}
\newcommand{\priority}[1]{{\color{red}[{\bf Priority:} #1]}}
\newcommand{\change}[1]{{\color{brown}#1}}
\newcommand{\cut}[1]
    {{\color{red}\sout{#1}}}	
	
%% CLEAR COMMENTS, no colours
% \newcommand{\notecn}[1]{}
% \newcommand{\changecn}[1]{#1}
% \newcommand{\notejs}[1]{}
% \newcommand{\changejs}[1]{#1}
% \newcommand{\notejc}[1]{}
% \newcommand{\changejc}[1]{#1}
% \newcommand{\todo}[1]{}
% \newcommand{\priority}[1]{}
% \newcommand{\change}[1]{#1}

%\headname{\begin{footnotesize}\color{red}Published in \emph{IEEE Access}, vol. 7, pp. 6562--6574, 2019. DOI: \href{https://doi.org/10.1109/ACCESS.2018.2887201}{10.1109/ACCESS.2018.2887201}\end{footnotesize}}

\setcounter{chapter}{4}
\numberwithin{section}{chapter}

\chapter{Towards an accountable Internet of Things\vspace{1mm}\newline \Large{A call for reviewability}}

%\todo{Sanity check}\notejs{Not sure if this will actually be included in the final -- but it will for archival sites}

% {\vspace{-6.8cm}
% \scriptsize
% \color{red} Cite as: ~Norval, C., Cobbe, J. and Singh, J. (2021). Towards an accountable Internet of Things: A call for `reviewability'. In \textit{Privacy by Design for the Internet of Things}. The Institution of Engineering and Technology.

% \vspace{6.5cm}
% }

\vspace{-7.5cm}\begin{scriptsize}\color{red}To appear: \hspace {.15cm} 
Norval C, Cobbe J, Singh J. Towards an accountable Internet of Things: A call for reviewability. In: \textit{Privacy by Design for the Internet of Things: Building accountability and security}; London, UK: IET; 2021.\end{scriptsize}
\vspace{6.8cm}

As the IoT becomes increasingly ubiquitous, concerns are being raised about how IoT systems are being built and deployed. 
Connected devices will generate vast quantities of data, which drive algorithmic systems and result in real-world consequences. 
Things will go wrong, and when they do, how do we identify what happened, why they happened, and who is responsible? 
Given the complexity of such systems, where do we even begin?

%This chapter outlines aspects of accountability as they relate to IoT, in the context of the increasingly interconnected nature of such systems. 
%We argue the need for effective logging mechanisms within IoT ecosystems as a means for enabling systems to be better designed, engineered and deployed in accordance with legal, regulatory, and societal concerns. 
%In all, this chapter indicates means in which IoT systems can be made to better support accountability, by enabling the relevant stakeholders in such an environment to better interrogate, challenge, and understand the connected environments that increasingly pervade our world.

This chapter outlines aspects of accountability as they relate to IoT, in the context of the increasingly interconnected and data-driven nature of such systems. 
Specifically, we argue the urgent need for mechanisms---legal, technical, and organisational---\textit{that facilitate the review of IoT systems}.
Such mechanisms work to support accountability by enabling the relevant stakeholders to better understand, assess, interrogate and challenge the connected environments that increasingly pervade our world.

%!TEX root = main.tex
\section{Introduction} \label{sec:intro}
%\todo{The emergent properties/behaviours point needs to be mentioned in sec 1 \& 2}\notejc{I've added a few mentions of it in both sections, let me know if it needs more}
% \begin{itemize}
%     \item Physical environments are becoming increasingly connected, but there is more to the IoT than just the devices
%     \item Lots of data from various sources feeding into algorithmic systems, which can result in consequential decisions/actions, near-instantaneously, and at scale
%     \item Things can and do go wrong, and we may find ourselves increasingly reliant on systems of which we have little understanding
%     \item There are increasing legal, regulatory, and societal concerns regarding the tech that surrounds us, and more demand for accountability
% \end{itemize}

Our physical environments are becoming increasingly interconnected, automated, and data-driven as grand visions of the Internet of Things (IoT) move closer to becoming realised. The concept of IoT involves a range of devices interacting and collaborating to achieve %common goals\cut{ and -- it doesn't have to be a common goal
 particular goals~\cite{iotvision}. %, with proponents promising more-informed, data-driven decisions. 
We already see many such examples of the IoT being deployed, including within homes and vehicles~\cite{dlodlo2016, mckee2017, xiong2012}, %(home security~\cite{dlodlo2016}, vehicle control~\cite{mckee2017, xiong2012}), 
and across towns and cities~\cite{asimakopoulou2011, boukerche2018, dlodlo2016, misbahuddin2015, saha2017, xiong2012}. %(traffic rerouting~\cite{misbahuddin2015, saha2017, xiong2012}, disaster management~\cite{asimakopoulou2011, boukerche2018, dlodlo2016}).
%We already see many such examples being deployed, including within homes and public spaces (security systems, smart heating, connected appliances)~\todo{cite}, and across wider areas such as towns and cities (dynamically routing traffic to ease congestion, smart street lighting, monitoring air quality)~\cite{zanella2014} \todo{more modern refs, e.g. from case study \S}. 
The claimed benefits afforded by so-called `smart' devices have led to considerable interest, with one estimate suggesting that the number of IoT devices could grow from 8 billion in 2019 to 41 billion by 2027~\cite{businessinsider2020}.
%\notejc{this is a very uncritical paragraph to begin with - it reads like we're just accepting these somewhat questionable claims about the IoT}\notejs{Adjust as you see fit. But we must begin with the fact this stuff is already being deployed, and lots of energy being put into it => probs in future}

There is, however, more to the IoT than just these `smart' devices. The IoT comprises a socio-technical ecosystem of components, systems, and organisations involving both technical and human elements. In these interconnected IoT environments \textbf{data flow drives everything}, in that the flow of data between, through, and across systems and organisations works  to both integrate them and deliver the overarching functionality~\cite{singh2016}. An IoT deployment will often include vast quantities of data coming from a range of local and external sources, including from sensor readings, online services, and user inputs, in addition to the data flows that occur across technical components (at various levels of technical abstraction) which drive particular functionalities. These data streams help to form wider algorithmic systems (which include degrees of automation, and may leverage the use of machine learning (ML) models), from which real-world consequences result.

These interactions introduce important socio-technical considerations regarding the wider contexts in which these IoT systems are built and deployed. As these ecosystems become ever more pervasive and complex, a range of devices, manufacturers, data sources, models, and online service providers may be involved in automated outcomes,  that can have impact at scale. Decisions made by people---vendors, developers, users, and others---about which systems and services to build, deploy and connect with, how these operate, their broader aims and incentives, and so forth, all serve to determine these data flows, with a direct impact on system functionality. Moreover, the complexity of these ecosystems means that systems can exhibit emergent properties and behaviours, acting in ways that are not explained by the features of their component parts when considered individually.
This raises important questions for how these systems should be %designed, developed, and deployed, 
governed, and what happens when things inevitably go wrong.

\subsection{Accountability}
%\todo{this sets the scene well - need to mention `supply-chain' in there somewhere}

Algorithmic and data-driven technologies have come under increasing scrutiny in recent years.
Reports of high-profile data misuse have made such systems increasingly the subject of legal and regulatory attention~\cite{diakopoulos2016, isaak2018}, leading to calls for \textit{more accountability} in the socio-technical systems which surround us.
Put simply, accountability involves apportioning responsibility for a particular occurrence and determining from and to whom any explanation for that occurrence is owed~\cite{bovens2006}. 
However, this can prove challenging in an IoT context. %~\cite{norval2019b}. 

Accountability is often discussed in relation to a single organisation, or in a systems-context as discrete systems, where the actors (individuals, organisations, etc.) involved are preknown or defined. %; e.g. concerning the actions undertaken by a particular organisation.
However, the various components of an IoT ecosystem---including sensors, actuators, hubs, mobile devices, software tools, cloud services, and so on---typically do not operate discretely or in isolation. Rather, they are employed as part of a \textit{system-of-systems}~\cite{singh2019}.
That is, these systems involve interactions with others, forming an assemblage of socio-technical systems~\cite{kitchin2014}. In this way, they are increasingly part of and reliant upon a data-driven supply chain of other (modular) systems, joined together to bring overarching functionality.
We already see many such systems-of-systems which are composed of interconnected components -- the popularity of cloud \textit{([anything]-as-a-service)}, which provide underpinning and supporting functionality on demand, is a case in point~\cite{cobbe2020}.

To illustrate, a given ecosystem may involve a range of devices streaming data to cloud-based ML services to produce new data outputs (a classification) which drives an automated decision. 
The decision itself then forms a (data) input to another system, which, in turn, results in that system producing an output (actuation command) for a device that works to create a physical interaction with real-world objects or people.
{An example of such an arrangement, for instance, might be a smart home management system, which receives inputs from various systems (e.g. weather forecast, temperature sensor, traffic congestion), and works to control various aspects of the home (thermostat, windows, alarm clocks, etc.).}
%\todo{tweak in line with DecProv caption. Smart home?}
%\changecn{This may be, for example, an array of sensors feeding into a weather prediction service, where that output is then fed into a building's temperature control system.}\notejs{This threw me off. Unclear and needs far better integration. Even if going as an example, very abstract - -what's the actuatioN?}\notecn{Tried to better integrate. I did just want an example to illustrate things a bit less abstractly. Hopefully it's a bit better now? I know not interesting, but we're just trying to illustrate the concept of data supply chains}
In practice, an IoT deployment may involve several of these interconnected systems and their data supply chains, encapsulating sensor readings, manual user inputs, the contents of web forms, inputs to or outputs from machine learned models, batch transfers of datasets, actuation commands, database queries, and more.

The {potential} complexity of these ecosystems {(which may comprise several such supply chains)} gives rise to significant accountability challenges, as it becomes increasingly difficult for those who build, deploy, use, audit, oversee or are otherwise affected by these systems to fully understand their workings.
It may not always be clear (i) why certain actions occur, (ii) where the information driving particular decisions, actions or functionality originates, and, in some cases, (iii) even which entities are involved. Things can---and will---go wrong, and we may find ourselves increasingly dependent on opaque, `black box' systems of which we have little oversight. The potential for emergent properties and behaviours in these complex environments further reduces the ability to understand what happened and why.

This chapter explores how greater accountability might be facilitated in the increasingly complex IoT deployments that surround us, arguing that implementing technical and organisational measures to facilitate the holistic review of IoT ecosystems is an important way forward.
%We argue that means that facilitating the review of IoT ecosystems is an important first-step towards greater accountability regarding the increasingly complex IoT systems that surround us.
We begin by elaborating on the challenges that complex systems-of-systems pose to accountability, before outlining the need for \text{reviewability} in such deployments.
Next, we discuss how reviewability can assist with legal accountability, before setting out some potential benefits for developers and other parties. %improving the accountability of the organisations and individuals responsible for them, and assisting them in meeting their obligations. 

We then set out some measures that may help implement reviewability, focusing particularly on \textit{decision provenance} which concerns recording information regarding decisions, actions and their flow on effects. Finally, we explore a `smart city' case study that illustrates how reviewability can assist complex IoT  ecosystems.
In all, we aim to draw more attention to accountability concerns in complex systems-of-systems, and argue that more needs to be done in order to facilitate transparency, reviewability, and therefore accountability regarding the IoT.
%!TEX root = main.tex
\vspace{-0.2cm}
\section{The need for reviewability} \label{sec:reviewability}
%\todo{The emergent properties/behaviours point needs to be mentioned in sec 1 \& 2}

\vspace{-0.15cm}

The IoT operates as part of a broader socio-technical ecosystem. As IoT deployments become increasingly automated and consequential, it becomes ever more important that they are designed with accountability in mind. In the event of failure or harm, a natural first step is to look at what went wrong and why. In determining legal compliance, it is important to understand how systems are and have been operating. Yet the complexity, opacity, interconnectedness, and modularity of these ecosystems pose particular challenges for accountability~\cite{singh2018}. Complex data flows across legal, technical, and organisational boundaries can produce emergent behaviours and can affect systems in unpredictable ways~\cite{singh2016,singh2019}. An unexpected behaviour or failure in one system or device can propagate {throughout the} ecosystem, with potentially serious consequences.

There has been some consideration by the technical community of issues of accountability, though these tend to focus on particular technical methods, components, or systems (see \S\ref{tech}). %\todo{add refs, or forward ref to tech implementation sec}.
Considering accountability by focusing only on particular technical aspects or components of the IoT 
%Focusing only on the technical accountability of particular systems and devices 
misses the bigger picture of their context, contingencies, dependencies, and of the organisational and human processes around their design, deployment, and use. 
We therefore argue here for a more holistic view of transparency and accountability in these complex socio-technical ecosystems. We term this approach \textbf{reviewability}. This encompasses a targeted form of transparency involving technical and organisational logging and record-keeping mechanisms. The purpose is to expose the information necessary to review and assess the functioning and legal compliance of socio-technical systems in a meaningful way and to support accountability (and redress, where appropriate). %, redress %\todo{split into two sentences or simplify}.

\subsection{Challenges to accountability in socio-technical ecosystems}

%\todo{1. P4: The authors might also want to mention that cross-system dataflow tends to create issues of contextual integrity, i.e. loss of information regarding the context in which the data can be made sense of appropriately, or regarding the norms of data uses associated with such data.}

As we have discussed, an IoT ecosystem is driven by data flows and may involve data moving between any number of different components and entities. As data passes from one entity or component to another (which can involve crossing  a technical  or organisational boundary), the data flows that drive these interconnected systems will often become invisible or opaque~\cite{singh2019, cobbe2020}. For example, it may not be clear to a developer how an external AI as a Service (AIaaS) model\footnote{This is where a machine learning model is offered as a service, and can be applied by customers to their data on-demand, typically as a request-response interaction. Example services include those for detecting objects, faces, converting text-to-speech, etc.~\cite{javadi2020}.} actually works or how reliable its predictions are. 
Similarly, it may not always be apparent where data provided by online services comes from, or how sensors are calculating and pre-processing data before feeding it to the next component. 
{This loss of contextual integrity~\cite{nissenbaum2004} risks the inadvertent misuse of that data, perhaps losing crucial information regarding how it should be used, information about bias{es} or sampling errors, or other potential problems that could continue to propagate throughout the wider ecosystem.}

Though there are methods for logging and ongoing research efforts on `explaining' technical components, their benefit may be limited in terms of improving accountability in an IoT context (\S\ref{tech}). 
As we will discuss, there is a need for tracking data as it flows between systems and across boundaries, given the interconnected nature of the IoT~\cite{singh2016b}. Moreover, even where  records of system behaviour (including data flow) are collected, the usability and utility of this information remains a recognised challenge~\cite{crabtree2018,norval2019}. %~\cite{bachour2015, crabtree2018, norval2019, schreiber2017} \notejs{give fewer refs in this section, to be consistent with the rest of this section (more in the tech section is fine)}. 

Having multiple components involved in a given deployment also increases the number of points of possible failure, and the potential for unpredictable emergent properties and behaviour. This increases the difficulty of determining where issues are occurring. As data flows through an ecosystem, decisions and actions taken by one component might propagate widely, which makes it difficult to trace the source (organisational or technical) and all the consequential (flow-on) effects of an issue. For example, a reading from a faulty sensor could lead to cascading knock-on effects causing significant disruption throughout an IoT deployment, though the relationship of the fault to these consequential effects may not be readily evident given the gaps in time and space~\cite{singh2018,singh2019}. Developers will often lack the means for `seeing' inside the various systems operated by others that comprise an IoT ecosystem. And although developers might have an understanding of the organisational processes in the development and deployment of their own systems, they are unlikely to have a similar understanding in relation to the components and systems on which their product will depend.

In all, the general opacity surrounding the interconnections between systems, of the technical and organisational processes within and around these systems, and with the general lack of technical means for tracking data flow at such a scale, poses a challenge for improving IoT accountability. Due to these factors, it can be difficult to discern the technical components involved in a given decision or action, why a failure has occurred, and who is responsible when failures do occur~\cite{singh2016b, pasquier2017}. While technical mechanisms for recording the flow of data between and across systems are necessary to identify where failures originated and which entities are responsible, there is a need to make this and other information, from right throughout the entire system lifecycle, {meaningfully available} across a broad range of stakeholders~\cite{cobbe2021}.

Importantly, providing technical ``transparency" on its own is not necessarily sufficient to provide the information required to understand the functioning of complex socio-technical ecosystems. We therefore continue this chapter with a focus on the \textit{reviewability} of IoT systems, components, and organisational processes -- providing targeted transparency and accountability mechanisms that allow for a holistic view to be taken of the socio-technical ecosystem as a whole.
Auditors and others assessing systems can then `hone in' on  specific systems, components, organisations and their practices where further investigation is required.

\vspace{-0.25cm}

%\notejs{I think in the text from the whole paper to this point we need to make clearer/reinforce the role of data and the exchanges that drives systems. We do say it, but probs only only be clear to those in the know. Even scope to say organisations exchange data too (hence DP regulations), but this exacerbated by the IoT.}\notejc{tried to emphasise data flow a bit more}
\subsection{Reviewability}

In view of these challenges to accountability in complex IoT ecosystems, we argue for \textit{``reviewability"} as an approach for shedding light on socio-technical processes. 

%\notecn{Too many boxes?}\notejs{this mirrors the text.. waht do we think -- is it OK for the same text to appear twice?}\notejc{I don't know what you mean by the same text? It's not repeated anywhere else (I don't think anyway). What's in this box was originally the first paragraph of this subsection to set out what we mean by reviewability in one para before getting into a bit more detail about why and what it is. I'm not sure it needs to be in a box}\notejs{It was both in the text here AND in a box. THInk only one place is better. I don't mind -- happy with box (breaks up text) as long as it shows up in the right place}

%\begin{figure}[!t]
\begin{boxes}
{\boxhead{Reviewability}}
{Reviewability involves {systematically} implementing {comprehensive} technical and organisational {transparency} mechanisms that allow the design, deployment, and functioning of socio-technical systems and processes to be reviewed as a whole~\cite{cobbe2021}. Reviewable systems are those that are designed and operate in such a way as to record and expose (though means such as, for example, record-keeping and logging) the {contextually appropriate} information necessary {to allow technical systems and organisational processes} %\notejs{@Chris: i put back tech/org processes. I might be missing a reason why you think that should be cut?}\notecn{Not specifically, it was just changed to more closely align with the definition in Jen's paper} 
to be comprehensively interrogated and assessed for legal compliance, that they are operating appropriately, and so on. This targeted, holistic form of transparency supports meaningful accountability of systems, processes, and organisations to whomever is relevant in a given context (auditors, investigators, regulators, users, other developers, and so on). 
}
\end{boxes}
%\end{figure}

% Submitted version
% Reviewability involves implementing comprehensive technical and organisational mechanisms that allow the design, deployment, and functioning of socio-technical systems and processes to be reviewed as a whole. Reviewable systems, therefore, are those that are designed and operate in such a way as to record and expose, e.g. though means such as record-keeping and logging, the information necessary to allow technical systems and organisational processes to be comprehensively interrogated and assessed for legal compliance, whether they are operating within expected parameters, and so on. This targeted, holistic form of transparency supports meaningful accountability of systems, processes, and organisations to whomever is relevant in a given context (auditors, investigators, regulators, users, other developers, and so on). 

% Jen version
% Reviewability as a general concept involves technical and organisational record-keeping and logging mechanisms that expose the contextually appropriate information needed to assess algorithmic systems, their context, and their outputs for legal compliance, whether they are functioning within expected or desired parameters, or for any other form of assessment relevant to various accountability relationships.

From a legal point of view, accountability is often tied to notions of responsibility, liability, and transparency -- indeed, transparency is often a regulatory requirement for identifying responsibility or liability and thus facilitating accountability, as we will discuss in \S\ref{sec:legal}. However, from a legal perspective, `transparency' does not necessarily mean full transparency over the internal workings of a system, nor is it limited only to technical components and processes. Instead, the transparency required in law often involves information about the entities involved, high-level information about what is happening with data or about what systems are doing (rather than necessarily the specifics of how they function), and information about the risks of using those systems \cite{singh2018}. 

In other words, legal mechanisms for accountability depend upon the ability of stakeholders to meaningfully review technical and organisational systems and processes---partially or as a whole---in order to determine which person or organisation is responsible for a particular system, device, or process, its effects, and from (and to) whom an explanation or redress is owed~\cite{singh2018}.
This may or may not involve exploring the inner workings of particular technologies or algorithms, as tends to be the focus of the technical research community. It could equally involve examining the broader socio-technical processes in and around a technical system itself. Indeed, there is debate about the degree to which exposing the details of code and algorithmic models actually helps with accountability~\cite{kroll2017,cobbe2021}. Therefore, instead of exposing the inner workings of technical systems and devices, achieving reviewability of IoT ecosystems requires (i) technical and organisational mechanisms for making transparent the connections and data flows across these entire ecosystems, as well as (ii) information relating to decisions made in design, deployment, operation, and during investigations, so as to indicate the context in which they are operating, their effects, and the entities involved (of which there may be a number). {To support meaningful accountability, the information recorded about these systems and processes should be \textit{contextually appropriate}~\cite{cobbe2021} -- that is, information which is \textit{relevant} to the kinds of accountability involved; \textit{accurate}, in that is correct, complete, and representative; \textit{proportionate} to the level of transparency required; and \textit{comprehensible} by those to whom an account will likely be owed.}

In practice, the information provided by reviewable systems will include that %in many cases be
 %considered `high-level' in nature, 
 indicating who is involved, the nature of their role, how data is being processed, and how data flows between systems for each component in the IoT ecosystem.
%The general idea is conceptually similar to the tracking of physical items through product supply chains, which helps manufacturers gain a better understanding of the provenance of their products and of the materials therein \cite{new2010}. 
%This allows factories and warehouses to respond and be accountable to manufacturers, which in turn allows manufacturers to be accountable to regulators and consumers. 
In an IoT context, reviewability could mean device vendors and service providers keeping records of, for example, the systems they are connecting to and interacting with  (in terms of input and output), what choices were made about which kinds of data to process and how it should be processed, information about assessments for security, data protection, and other legal obligations, and information about the design, training, and testing of models (where applicable). At its most basic, this kind of information may indicate that Organisation B used an online service (e.g. AIaaS) provided by Organisation C, and that the response from that prediction was then fed into a service provided by Organisation D. But an analysis of even this `high-level' information could also provide an indication of where the source (either technical or otherwise) of a particular issue may lie. In this sense, such information can make it easier to review what went wrong (particularly for non-experts), prompting more targeted investigations involving the inner workings of particular components within the IoT ecosystem once they have been identified, of which reviewability mechanisms that provide more detailed and granular information will facilitate. 

Of course, implementing reviewability so to support the transparency and accountability of socio-technical systems introduces challenges of its own, given that what is meaningful for one stakeholder, such as a technical auditor, may be very different to that of another, such as a user. There is a need to ensure that the information provided by various transparency mechanisms {is contextually appropriate such that it} is useful for and can be interpreted by {the likely} audience~\cite{cobbe2021}. This might involve, for example, different ways of presenting or processing various kinds of information for a broad range of stakeholders -- as we discuss below. 

While transparency as a general principle is important, it cannot in and of itself solve problems. Organisations need to think more about targeted transparency to support accountability -- from the requirements stage, all the way up to deployment, operation, and beyond. Regardless of the approach to enabling meaningful transparency, reviewability should be a core consideration for those involved in the development and operation of IoT ecosystems. In the following sections, we explore some legal and technical ways forward that could help toward this aim.

%\notejs{this covers a lot of the key points -- will need to be reconciled with the text in the latter sections}

%!TEX root = main.tex

%\section{Data protection and the IoT} 
\section{The legal dimension} 
\label{sec:legal}
%\todo{@Jen - we'll need to cite this at the appropriate points -- brings up a lot of the liability challenges: \url{https://ieeexplore.ieee.org/document/7923814}}
% - While tech solutions may provide part of the puzzle, wider motivations may be a challenge. Legal approaches might play a role [Intro to GDPR]

%\begin{itemize}
%	\item current means inadequate
%	\item how reviewability may assist (useful for GDPR obligations to DPAs, need for oversight for liability, consumer standards, etc)
%	\item ways forward ??
%\end{itemize}

%The complex, interconnected nature of the IoT ecosystem poses various challenges for legal accountability. Broadly speaking, these challenges come from two sources: (1) the lack of visibility over data flows in interconnected environments, which makes it difficult to know where data has come from or where it is going to; and (2) the lack of visibility over the technical and organisational systems and processes of various entities within that ecosystem, which makes it difficult to assess their compliance with legal requirements and obligations.
The complex, interconnected and data-driven nature of the IoT ecosystem poses particular challenges for legal accountability ~\cite{millard2017, singh2018}. Broadly speaking, these challenges come from two sources: (1) the lack of visibility over data flows in interconnected environments, which makes it difficult to know where within a system and with whom a problem originates; and (2) the lack of visibility over the technical and organisational systems and processes of various entities within that ecosystem, which makes it difficult to assess their compliance with legal requirements and obligations. 
%\notejs{does (2) subsume (1)?}\notejc{does the next bit of text clarify it?}\notejs{yep. worked some of that into the above }
%These two issues together work to reduce the levels of accountability in these interconnected environments -- it will often be impossible to determine where a problem originates without knowing where data is flowing to and from; equally, simply knowing where data is flowing to and from doesn't provide information about how other component systems are functioning (or not).

From a legal point of view, each entity in an IoT ecosystem could potentially be accountable for the functioning of their systems and devices -- to customers and users; the vendors of other IoT devices and systems; to regulators and other oversight bodies; to law enforcement agencies; and to courts and dispute resolution bodies; among others. For each of these, `transparency' as a general concept is unlikely to provide much benefit, nor is opening up code or the internal workings of technical systems likely to assist in every case (although it will, of course, in some~\cite{elecevid})~\cite{cobbe2019}.
Rather, the information necessary to meet accountability and disclosure requirements (and obligations established in law) is more likely to relate i) to technical and organisational processes, ii) to data flows given the nature of the IoT, and iii) to understanding what happened, when, why, and with what effect~\cite{singh2018}.

For example, data protection law, such as the EU's General Data Protection Regulation (GDPR)~\cite{gdpr}, establishes accountability mechanisms and requirements for various actors involved in processing personal data (defined broadly in GDPR to include any data from which an individual can be identified, whether directly or indirectly, either from that data alone or when combined with other data\footnote{GDPR, Art. 4(1), recital 26}). IoT ecosystems will often involve a flow of personal data, and \textit{processing} means any operation or set of operations performed on that data.\footnote{GDPR, Art. 4(2)} Under GDPR, certain information is required to be provided by data controllers (the entities that determine the purposes and means of processing\footnote{GDPR, Art. 4(7)}) to data subjects (the individuals to whom personal data relates\footnote{GDPR, Art. 4(1)}) about the processing of their personal data. Data processors (who process personal data on behalf and under the instruction of data controllers\footnote{GDPR, Art. 4(8)}) are obliged to facilitate auditing of their processing by the relevant data controller. Data controllers can be obliged to provide information about processing to and facilitate auditing by supervisory authorities (national data protection regulators). Controllers will also need to know where personal data has been obtained from, under what circumstances, and on what conditions. 

Beyond data protection law, accountability requirements could also arise from a variety of other sources~\cite{millard2017} -- product safety law, for instance, could require device vendors to supply customers and oversight bodies with information. The adjudication and enforcement of contractual disputes could require disclosure of information about technical and organisational processes to counterparties and to courts. Criminal investigations could require the provision of information to law enforcement agencies.

In each case, approaching transparency through the lens of reviewability could assist in meeting these obligations, and help communicate legally relevant{, contextually appropriate} information to customers, users, technical partners, and the appropriate legal, regulatory, and other oversight authorities. Although reviewability could assist across various legal frameworks, we now explore three general ways in particular where a reviewability approach to accountability would be legally beneficial: compliance and obligation management; oversight and regulatory audit; and liability and legal investigation. It is worth noting that while we have delineated these benefits, they are interrelated. As much of the IoT ecosystem will involve personal data, we use the transparency and accountability requirements and obligations established in GDPR as examples throughout.

\subsection{Compliance and obligation management.} \label{compliance_ob_management} 

%\notejc{had a go at chopping this down}\notejs{I think this looks good. Think DPIAs need a short mention, as they're sometimes required, entails documentation, and mentioned below (without the acronym being introduced)}\notejc{added a bit}

As alluded to above, a reviewability approach to accountability---such as by using record-keeping, logging or other information capture mechanisms for technical and organisational processes---can assist those responsible for socio-technical ecosystems (such as the IoT) to comply with their legal and regulatory (and other) obligations~\cite{singh2015,singh2016b}. 

Where IoT ecosystems process personal data (perhaps, for example, containing names, user profiles/accounts, photographs or video of people, voice recordings, or any other information that can be linked to an individual), and therefore come within the remit of data protection law, GDPR would typically consider those involved in operating the various aspects of that ecosystem to be data controllers. This will therefore place upon them certain legal obligations, including making them responsible for meeting GDPR's requirements and obligations, and for demonstrating compliance.\footnote{GDPR, Art. 5} In this context, reviewability would not necessarily involve recording the personal data itself---a potentially significant data protection issue and privacy violation---but would involve metadata about organisational processes and technical processing. Organisations found to be in breach of the GDPR can face significant penalties, including warnings, fines of up to the greater of \euro{20m} or 4\% of annual global turnover, and bans on processing.\footnote{GDPR, Art. 83}

In the first instance, data controllers are obliged to take technical and organisational measures to ensure and be able to demonstrate compliance with GDPR.\footnote{GDPR, Art. 24} Record-keeping and logging would assist data controllers with implementing GDPR's data protection principles and with showing that they have taken steps to do so. For instance, the `purpose limitation' principle requires that personal data be collected for specific purposes and only processed in a way compatible with those purposes.\footnote{GDPR, Art. 5(1)(b)} The keeping of records, logs and other relevant information  would provide controllers with relevant information on the purposes for which personal data was collected, and helps ensure that the data is only processed accordingly. Implementing such measures would also assist with fulfilling data subject rights, such as those to obtain copies of the personal data being processed \footnote{GDPR, Art. 15} and, in some circumstances, to require its deletion.\footnote{GDPR, Art. 17} Moreover, maintaining records  of technical and organisational processes around the design and deployment of systems would help controllers demonstrate to supervisory authorities that processing is in fact taking place in line with GDPR's requirements and that rights have been fulfilled.

In advance of processing, data controllers will in many situations need to undertake a Data Protection Impact Assessment (DPIA).\footnote{GDPR, Art. 35} This consists of a broad assessment of the risks posed by the proposed processing to the rights and freedoms to data subjects and of the steps taken by the controller to mitigate those risks. Where the DPIA indicates a high risk, controllers are obliged to consult with the supervisory authority before proceeding and to provide them with the information contained in the DPIA.\footnote{GDPR, Art. 36} Reviewability can greatly assist with undertaking the kind of holistic analysis of processing and its risks required for DPIAs, by demonstrating how and whether any mitigating measures have actually been implemented, and with assessing on an ongoing basis whether processing is in fact being undertaken in line with the DPIA.

It is worth noting that controllers and processors are also already obliged in many circumstances to keep records of processing.\footnote{GDPR, Art. 30} Controllers should record, among other things, the purposes of processing, the categories of data subjects and of personal data, the categories of those to whom data will be disclosed, and a general description of technical and organisational security measures. Processors should record, among other things, the name and contact details of processors, the categories of processing, and a general description of their technical and organisational security measures. Through comprehensive logging and record-keeping of technical and organisational systems and processes, reviewability can help controllers and processors fulfil these obligations and others.

%\notejc{need to include this para somewhere}\notejs{Moved reactive bit to technical section, needs better integration though} 
%Reactive, event-based mechanisms could also automatically take actions to assist compliance and obligation management~\cite{singh2016b}. For instance, information about data flows could be used to trigger particular compliance operations; such as to automatically report data breaches to the relevant authorities; to screen and filter out data based on compliance criteria (e.g. data past an expiry date or use); to not act on inputs that may have come from, or through, an unreliable entity~\cite{singh2015}; or to automatically prevent data flows that are unexpected (in terms of a pre-defined management policy~\cite{singh2016b}). Such approaches may be particularly useful to IoT deployments given their often emergent nature.

Reviewability may also be of use in assisting compliance with future regulations; for example, the proposed ePrivacy Regulation~\cite{epriv}, which establishes requirements on the use of some non-personal data. They can similarly assist in managing a broader range of legal, regulatory, and other obligations (`soft law', codes of practice, and so on). For instance, where contractual obligations exist between entities, knowledge of their respective technical and organisational processes (and of the nature of data flow between parties) could make it possible to ensure and demonstrate that data is processed in a legally appropriate manner. And information on the selection of datasets, and on the sources and lineage of data used for analytics and ML, might assist with issues of unfairness and discrimination~\cite{wef2018}.

\subsection{Regulatory oversight and audit.}
\label{law:oversight}
Reviewability also has much potential to aid the auditing and oversight activities of regulators, given that reviewability entails generating detailed information for assisting oversight that may otherwise be unavailable. As noted previously, data controllers and processors are obliged to maintain extensive records of processing. Data controllers are also obliged to undertake various assessments (in relation to data protection by design, for example, or for implementing security measures, carrying out DPIAs, and so on), to use only processors who can comply with GDPR, and to consult with supervisory authorities in certain circumstances. Supervisory authorities are themselves empowered to audit data controllers, to inspect their systems, to access to their records, and to require from controllers any other information necessary to perform the supervisory authority's tasks, including to investigate controllers' compliance with GDPR.\footnote{GDPR, Arts. 57-58}

Engineering socio-technical systems to be reviewable through mechanisms for record-keeping, logging and otherwise capturing information about of technical and organisational processes would thus facilitate audit and oversight by supervisory authorities. Indeed, many of the potential legal benefits of reviewability discussed previously would also apply to regulatory oversight more generally. Moreover, recording information about data flows would allow data protection regulators to assess whether IoT ecosystems are legally compliant and match that which was described by controllers. More broadly, this information would help supervisory authorities to assess whether data controllers and processors are managing their relationship as required, have the appropriate data management (and other) practices in place, and have taken the appropriate actions in relation to any specific incidents such as data breaches.

%In the event of non-compliance, supervisory authorities---those responsible for enforcing the regulation, appointed by each member state---have means for holding data controllers to account.
%For assisting with this, the GDPR also contains powers for regulators to request certain information from data controllers.\footnote{GDPR art.58} 
%This includes a wide-ranging power to obtain any information from data controllers that is necessary for the performance of the regulator's tasks,\footnote{GDPR art.58(1)(a)} and includes the power to audit controllers.\footnote{GDPR art.58(1)(b)} 

%Beyond regulators and oversight bodies, the same methods are useful for data controllers to help in managing their own data supply chains.
%For example, such information assists an organisation in evaluating that online services that their IoT deployment engages with (e.g. through a data processor relationship) are meeting their contractual and data management obligations by handling data appropriately.

\subsection{Liability and legal investigation.} \label{subsec:legal_benefits-liability} 

%\todo{3. P11: The authors might want to cite Art 82(3) GDPR to further support their claim that reviewability may help organisations to demonstrate they are not at fault for an incident, and therefore can be exempt from joint liabilities.}

IoT systems have real-world impacts, including direct physical-world consequences given that IoT systems are people-facing and actuators can feature throughout. This means that questions of civil liability or criminal responsibility could arise where harm or injury is caused~\cite{millard2017, singh2018}. Reviewability could therefore be particularly helpful in assessing liability and undertaking legal or criminal investigations. Where harm is caused by a system failure, comprehensive records and logs would allow the circumstances, causal behaviours, factors, and consequences of the failure to be reviewed, determined, and liability established. In complex, interconnected environments, records and logs about data flows to, from, and between devices and systems could also help identify which system caused that harm. This, thereby, helps to identify the component (and entity) responsible and potentially liable for that harm, and can assist in holding them to account. This kind of reviewability information may even help absolve systems designers, operators or those otherwise involved from responsibility, by providing evidence demonstrating that the right decisions were made and actions were taken. {For example, controllers and processors are exempt from liability for damage caused by infringements of GDPR if they can prove that they were not responsible for the infringement (such as where a controller can demonstrate that their processor acted unlawfully, for instance).}\footnote{GDPR, Art. 82(3)}

As such, the information provided by implementing reviewability is useful for multiple actors in interconnected IoT ecosystems -- to vendors, in helping them manage their liability exposure; to users, in taking action and obtaining redress where harm has been caused by a product; to courts, in adjudicating claims in tort or contract; as well as to regulators and law enforcement as they investigate potential regulatory or criminal violations. Vendors and others could contractually require this kind of record-keeping and logging (of technical and organisational processes and data flows, both within organisations and across boundaries). This could give them a mechanism for monitoring how data is used, potentially in real-time, and enable action where data is being used in a way that is inappropriate or prohibited by law or by contract.

\section{The broader benefits of systems review}%\notejc{agreed, should change both - not sure what to, though} 
\label{sec:technical}

In an IoT context, there are clear benefits of reviewability, in addition to those legal as just discussed. 
In \S\ref{sec:reviewability}, we described how the complex, interconnected, data-driven, and multi-stakeholder nature of the IoT can result in transparency challenges at the technical level, making it difficult to determine how systems are designed, deployed, and functioning. The opacity of these complex systems-of-systems arrangements means that the information necessary to properly review their functioning may not be available, which can significantly hinder accountability processes.  %potentially significantly hindering accountability processes.
%Even those who have built and designed technical components may suffer from a lack of information to properly oversee their function in practice; the problem exacerbates when considering the broader interconnected and socio-technical information that would be required in such a context.
% As a result, developers and other technical parties may lack sufficient information to assess how these systems are functioning.

Information about the organisations, components, and data flows driving IoT ecosystems therefore has an important role to play in supporting review processes. That is, certain details and related information about IoT systems and components and their design, deployment, and operation---which includes workflows, business processes, and other organisational practices---will often be necessary for enabling proper oversight, audit, interrogation, and inspection. Moreover, the \textit{ongoing} review of the socio-technical systems is important. This is because  the complexity,  interconnectedness, and `long-lived' nature of IoT ecosystems makes it more likely that they will exhibit emergent properties and behaviours, potentially giving rise to unforeseen or unforeseeable problems.

%In arguing the need for such, we now explore how technical information can assist different actors, and indicate the potential for provenance-based approaches as one method showing promise.

%We argue for more attention on technical approaches to support and facilitate review. 
%In practice, this entails mechanisms that capture information about components, systems and their interactions comprising IoT systems.  We now explore some of the benefits of such technical information, and indicate the potential for provenance-based approaches as one method that assists.

%\subsection{Benefits: technical information for reviews} %benefits

%Information of the systems and components driving the IoT has an important role in supporting review processes.
%This is because such information facilitating audit and interrogation, which is crucial for ensuring systems are designed and behave appropriately, and also for unpacking issues when they arise. 
%Given the IoT entails an ecosystem comprising a variety of stakeholders, there will be a range of review processes, that depends on the particular actor and their aims. It follows that having more details and visibility over the technical aspects of the IoT will benefit different actors in different ways. We illustrate this in general terms, considering some classes of key stakeholders in the paragraphs below.

In short, information about the design, deployment, and functioning of the socio-technical systems that comprise the IoT facilitates their review. This includes whether they are designed and behaving appropriately, and for issues to be unpacked and consequences dealt with when they arise. This brings additional benefits to  the legal aspects described in \S\ref{sec:legal}, for a range of actors in an IoT ecosystem.

%\notejs{this next bit is framed around actor... another approach is to breakdown into -- post hoc investigation, proactives resposne (alerts/preventtion) then talk about how it would help each stakeholder. Thoughts? Does this fit with the legal articulation?}\notejc{I think that probably makes sense and more closely reflects how the legal section is structured - by what kinds of benefit it gives you, rather than by who gets what kind of benefit. I think an ex ante / ex post thing could work. Or maybe a design time / run time thing? Or I suppose those two could combine in some way}\notejs{I left it as it is, framed aroudn the actor. This is because there's not really a lot to say generally about each -- the benefits ultimately are actor-specific. If we tried to frame benefits around the functionality (i.e. what they enable) it then relates to the specifics of the supporting technology/methods... we're taking a specific approach towards capturing information\\ Welcome to try alternatives.}

%\subsection{Assisting investigation and audit (ex-post)} %ex-post/ex-ante is too jargony for this audience

%\vspace{.8mm}\noindent \textit{Technology developers and operators:~} 
%\vspace{.8mm}\noindent \textit{Helps technologists better build, operate and manage systems:~} 
\subsection{Building, operating, and managing systems}
Implementing technical and organisational measures for logging, record-keeping or otherwise capturing information of systems increases their reviewability. This brings a range of benefits for technologists, who have a clear interest in overseeing the systems they build and operate.
This is because details on how systems are designed, deployed, function, behave, interact, operate, and so forth provide information relevant for testing, monitoring, maintaining and improving the quality of those systems, while helping them to meet and manage their responsibilities and obligations. 
%In an IoT context, more technical details about systems---both at design and run-time---regarding the components of the systems they operate, and also details of  their interactions with other systems improves an organisation's capacity for such undertakings. 

Specifically, greater visibility over the technical and organisational aspects of systems, and their interconnections with others, assists technologists with investigations, enhancing their ability to review errors and systems failures when they (inevitably) occur. That is, such information helps support processes of repair and debugging. Moreover, information about run-time behaviour also enables more pro-active steps to be taken, where details of system operation can allow the  identification and mitigation of certain issues in advance of them becoming problematic -- for example, by enabling the identification of abnormal deviations in system behaviour, or through run-time reaction mechanisms, alerts, or automatic countermeasures that are triggered where unexpected events or interactions occur. This is particularly important given that, as discussed previously, IoT systems can exhibit emergent properties and behaviours. Further, in the IoT, components have the potential to be used or reused in ways that were not envisaged by the original designers; a temperature sensor in a building designed for climate control could suddenly be used to influence health wearables, for instance. It follows that the ability for technologists to undertake ongoing review will be crucial for ensuring appropriate system behaviour, while providing insight into (any) emerging  properties.

%The ability to uncover such is cru\notejs{..}

%Monitoring system behaviour, and `bounding' (or constraining) particular outputs will be important, given that IoT arrangements lend themselves to emergent properties and behaivour. This means that information that supports ongoign review processes will be crucial for this.

%Exposing the supply chain of a particular decision or outcome can help in identifying aspects such as the steps behind the leakage of personal or otherwise sensitive data, and in tracking the cascading consequences of a particular (perhaps erroneous) event.

%\vspace{.8mm}\noindent \textit{Oversight bodies:~} 
%\vspace{.8mm}\noindent \textit{Facilitating oversight activities} 
\subsection{Facilitating oversight activities}
%\notejc{I took out the legal/regulatory oversight bit since it's discussed in 3.2, but I'm not sure that leaves us with much in this section that isn't covered by the previous one. If there's anything that's not included in 4.1, maybe we should move it to 4.1 and cut the rest of this?} Think leave it - it's a clear summation that reinforces the message
As we have described, the lack of technical information in an IoT context has the propensity to hinder investigations.
While reviewability promises benefits for legal and regulatory oversight bodies (see \S\ref{law:oversight}), information about technical and organisational systems can also assist the activities of other overseers, such as civil society organisations, trade-sector groups, or those internal to a technology firm. This is by giving (i) information on the nature of systems and their behaviour; and (ii) insight into the practices of their designers and operators. 

In this way, information on systems  helps accountability processes, by providing evidence or lines for investigation that can assist overseers in determining whether the technology-related actions and measures taken were appropriate (thereby building the foundations for recourse where they are not). This can include information about design and development, as well as run-time systems behaviour,{ and details of the reviews and audits undertaken}. In addition to supporting {(ex-post)} investigation, technical processes that produce streams of data on system operation also paves the way for more active and responsive oversight regimes. This is where certain system behaviours or events could trigger alerts, warnings or otherwise bring certain information to an overseer's attention, and with timely action, may serve to mitigate the potential severity of outcomes.  

%\vspace{.8mm}\noindent \textit{Enabling more informed users:~}  %\notej{I think we should keep the user aspect -- the editors love HCI, and it does tie in with the rights discussion ... if worried can make explicit somewhere that burdening a user isn't the goal there can be info assym benefits too}
\subsection{Better informing users}
The information derived from  IoT systems %, including records and logs of system design, deployment, and operation 
can potentially also assist end-users by helping them to make better-informed decisions about whether or how they interact with a given deployment. Although overloading users with information and options is not a desirable or useful outcome~\cite{stohl2016}, there may be  opportunities for a greater degree of user empowerment. If a user could meaningfully and reliably know in advance that engaging with (having their data flow into) a particular system could, for example, (i) result in certain decisions being made that may have particular undesired consequences; or (ii) involve data flowing to an undesired entity (such as a certain advertisement network), then the user could take reasonable steps to avoid engaging with such a system~\cite{singh2019}. 
Proactive measures are again possible, such as where user policies might work to constrain particular information flows~\cite{singh2016b}, or to keep users informed of any change in circumstance, thereby enabling them (manually or perhaps automatically) to respond. 
One can imagine, for instance, users being automatically told of changes in policy or system operation -- for example, where the records of data flows indicate an online service suddenly engaging a new advertising network.

%Such measures, driven by provenance information, could allow users to take action, e.g. exercise their rights, complain to a regulator, stop using the service.

%Details of technical systems can also assist oversight bodies---be they regulatory, civil society, or even internal agency---by facilitating their audit activities. This is by giving (i) information on the nature of systems and their behaviour; and (ii) insight into the practices of their designers and operators. 
%This helps accountability processes, providing evidence or lines of interrogation that can assist overseers in determining whether the actions and measures were appropriate, and thereby building the foundations for recourse where they are not.

%-enabling developers, designers and operators to audit/investigat their own systems, find problems, audit etc
%-enabling oversight agencies to audit, inspect, interrogate
%-enabling end-users to be better informed, potentially enabling empowerment
%

%!TEX root = main.tex

\section{Technical mechanisms for supporting reviewability}
\label{tech}
%\todo{@Chris - add crabtree logging ref}%\notejs{@Chris you had a citation for Crabtree re logging -- plz work this in somewhee}
%Indeed, prior research has considered audit logging~\cite{crabtree2018} and provenance~\cite{norval2019, wang2018} for use within IoT ecosystems.

%Note that the benefits of reviewability extend beyond standard, existing software design processes. Rather,  enabling ongoing system review is particularly important in an IoT context, given the complex ecosystem can lead to emergent behaviours and properties, perhaps designed for, nor previously envisaged. 

Technology has a key role to play in supporting reviewability, as it provides the means to capture or otherwise produce information or records about systems,  both about their design and during operation. As we have argued, reviews are assisted through mechanisms that can enable more (relevant) transparency over systems.  %However, so far, there has been little attention on technical approaches for supporting and facilitating review, in a manner appropriate for the broader visions of the IoT. In practice, this is beyond the information captured as part of contemporary systems design and operation (i.e. standard logging), but rather requires mechanisms for capturing information about components, systems, and their interactions (data flows) that comprise IoT systems.  

Though the idea of capturing information about systems is nothing new, much of the focus so far has been on the design and engineering phases. Logging and debugging frameworks, for example, are generally aimed at a technical audience, particularly those developing or administering systems. These tend to focus narrowly on particular technical specifics. There has been less consideration of recording information (i) to support broader accountability regimes (of which, we argue, reviewability enables), and (ii) in a manner appropriate for the broad vision for IoT.

That said, accountability is an area of increasing attention by the technical research community (e.g. see~\cite{crabtree2018,singh2018}). `Accountability' is a term that gained traction, particularly in the machine learning community where the focus has predominately been on the `explainability' of machine learning models and decisions. While such work is important---perhaps particularly for machine learning (engineering/testing) processes~\cite{wellerfat}---a model doesn't operate in isolation. Rather, it exists as part of a broader socio-technical system~\cite{singh2016c,cobbe2021}.  This broader aspect requires consideration, and work in the community is beginning to expand and recognise this~\cite{cobbe2021}. One example is \textit{`datasheets for datasets'}~\cite{gebru2018}, which involves capturing metadata that describes the nature of datasets, including how the data was collated, pre-processed, the legal basis for its processing (e.g. consent), etc.
{Similarly, \textit{`model cards'}~\cite{mitchell2019} record various details about trained models, such as how they perform across a variety of different conditions and\slash or demographic groups, alongside the context in which they are intended to be used.}
The idea is to provide more information, for instance, to enable a fuller interrogation of model design (e.g. how it was trained), help inform engineers and investigators, for example, as to whether the dataset or model is (or was) appropriate for a particular use, give information about usage restrictions (e.g. consent), and so on.

Considering the broader systems context is important in the IoT; IoT ecosystems are complex, interconnected, dynamic, multi-actor, socio-technical environments, which are long-lived and can potentially elicit emergent behaviours -- those not designed for or previously envisaged. 
As we have discussed, greater visibility over the interconnections and assemblages of systems is important for increasing accountability. 
Therefore, a natural starting point for reviewability is technologies which assist in uncovering or mapping out the (ongoing) nature of IoT ecosystems. This  requires mechanisms for capturing information about components, systems, and their interactions (data flows) that comprise IoT systems. 

Particularly important in an IoT context are the ongoing interactions (data exchanges) with other technical components, as these provide a foundation for both (i) describing system behaviour (the `how' and `why') and (ii) the actors involved (the `who').  However mapping this represents a particular challenge given that while the IoT is data-driven, the visibility over the movement of data currently tends to be limited to particular scopes; it is often difficult to trace what happens to data once it moves across a technical (e.g. between\slash across different software or devices) or administrative (to another organisation, department, etc.) boundary. %Moreover, long-liven nature of IoT systems requires capacity to review systems behaviour over time, given the complex ecosystem can lead to emergent behaviours and properties -- perhaps not designed for, nor previously envisaged~\todo{cite}.

Towards such a mapping, provenance mechanisms show real potential~\cite{aldeco2010, pasquier2017, singh2015, singh2016b, tan2013}. 
\textit{Provenance} concerns capturing information describing data: it can involve recording the data's lineage, including where it came from, where it moves to, and the associated dependencies, contexts (e.g. those environmental and computational), and processing steps~\cite{carata2014}.\footnote{See ~\cite{carata2014, herschel2017} for more details on data provenance.}

To reiterate, the IoT is data-driven, where functionality is brought about through the interactions---the exchanges of data---between systems components. As such, provenance methods can be useful for reviewability as they entail taking a `follow the data' approach -- whereby information regarding the data flows in the IoT deployment can indicate how systems behave, what led (or leads) to particular occurrences or consequences, as well as the components (and therefore the actors) involved~\cite{singh2016b,pasquier2017}.
Data regarding the information flows driving systems also provides the foundation paving the way for new accountability-related tools, methods, and analysis~\cite{singh2018}.
%We argue that one method forward, certainly relatd but beyond that, are mechanisms that are provenance related.
%Provenance entails capturing the flow of data within/across systems.
%In terms fo complex, multi-stakeholder environs, we look at decprov

In line with this, we have developed the concept of \textit{decision provenance}~\cite{singh2019}, which can help in mapping out complex ecosystems. We argue this has real potential in the IoT space, accounting for its interconnectedness, and the role of automation (including the use of ML) in such environments. It works by recording what (and how) data moves from one system (device, model, service, etc.) to another.
We now present an overview of decision provenance to indicate the potential for technical methods which can assist review, and support accountability regimes more generally.  

%\vspace{-1mm}

\subsection{Decision Provenance: Exposing the decision pipelines} \label{subsec:decprov}

% \begin{itemize}
%     \item Provenance concerns capturing the lineage, dependencies, contexts, and processing steps of data [examples, use cases]
%     \item Decision provenance concerns the use of provenance mechanisms to assist accountability considerations in algorithmic systems [definition]
%     \item Decision provenance helps expose the decision pipelines in order to make visible the inputs and cascading consequences of any decision or action
% \end{itemize}

Provenance is an active area of research~\cite{herschel2017}, and is commonly applied within research contexts to assist in reproducibility by recording the data, workflows and computation of scientific processes~\cite{herschel2017, davidson2008}. 
The potential for provenance to assist compliance with specific information management obligations has previously been considered~\cite{aldeco2010, singh2015, tan2013, singh2016b, pasquier2017}, though these often focus on a particular technical aspect, be it representation or capture. 
However, the use of data provenance methods for general systems-related accountability concerns represents an emerging area warranting further consideration.
%We now outline one such potential approach for using provenance techniques for facilitating greater accountability in an IoT context.

%Given (i) the concerns around accountability for automated, algorithmic, and data-driven systems (of which, many exist within the IoT), (ii) the increasingly interconnected nature of IoT deployments, and
%(iii) the potential of provenance methods in this space, we outline \textit{decision provenance}~\cite{singh2019} as one potential way forward.

Decision provenance is a response to (i) the concerns around accountability for automated, algorithmic, and data-driven systems (which naturally comprise the IoT), (ii) the increasingly interconnected nature of technical deployments, where functionality is driven through data exchanges, (iii) the current lack of visibility over data as it moves across technical and administrative boundaries, and
(iv) the potential of provenance methods in this space. %, we outline \textit{decision provenance}~\cite{singh2019} as one potential way forward.

Specifically, decision provenance concerns recording information about the data flowing throughout a system, as relevant for accountability concerns. It involves capturing details and metadata relating to data, and of the system components through which data moves. This can include how data was processed and used, who the data comes from or goes to (by virtue of the components involved), and other appropriate contextual information, such as system configurations, business processes, workflow state, etc.

%\begin{figure}[!t]
\begin{boxes}
{\boxhead{Decision Provenance}}
{Decision provenance concerns the use and means for provenance mechanisms to assist accountability considerations in algorithmic systems. 
Specifically, decision provenance involves (i) providing information on the nature, contexts and processing of the data flows and interconnections leading up to a decision or action, and the flow-on effects; and (ii) also how such information can be leveraged for better system design, inspection, validation and operational (run-time) behaviour. 
In this way, decision provenance helps expose the \textit{decision pipelines} in order to make visible the nature of the inputs to and cascading consequences of any decision or action (at design or run-time), alongside the entities involved, systems-wide.
 
The broad aim is to help increase levels of accountability, both by providing the information and evidence for investigation, questioning, and recourse, and by providing information which can be used to proactively take steps towards reducing and mitigating risks and concerns, and facilitating legal compliance and user empowerment. For more details on decision provenance, see \cite{singh2019}.
%\todo{@chris - diagram needed} \notecn{I couldn't add this into the box (it wouldn't render). Added a first version outside.} 

%\vspace{1mm}
\begin{center}
\noindent\includegraphics[width=0.8\columnwidth]{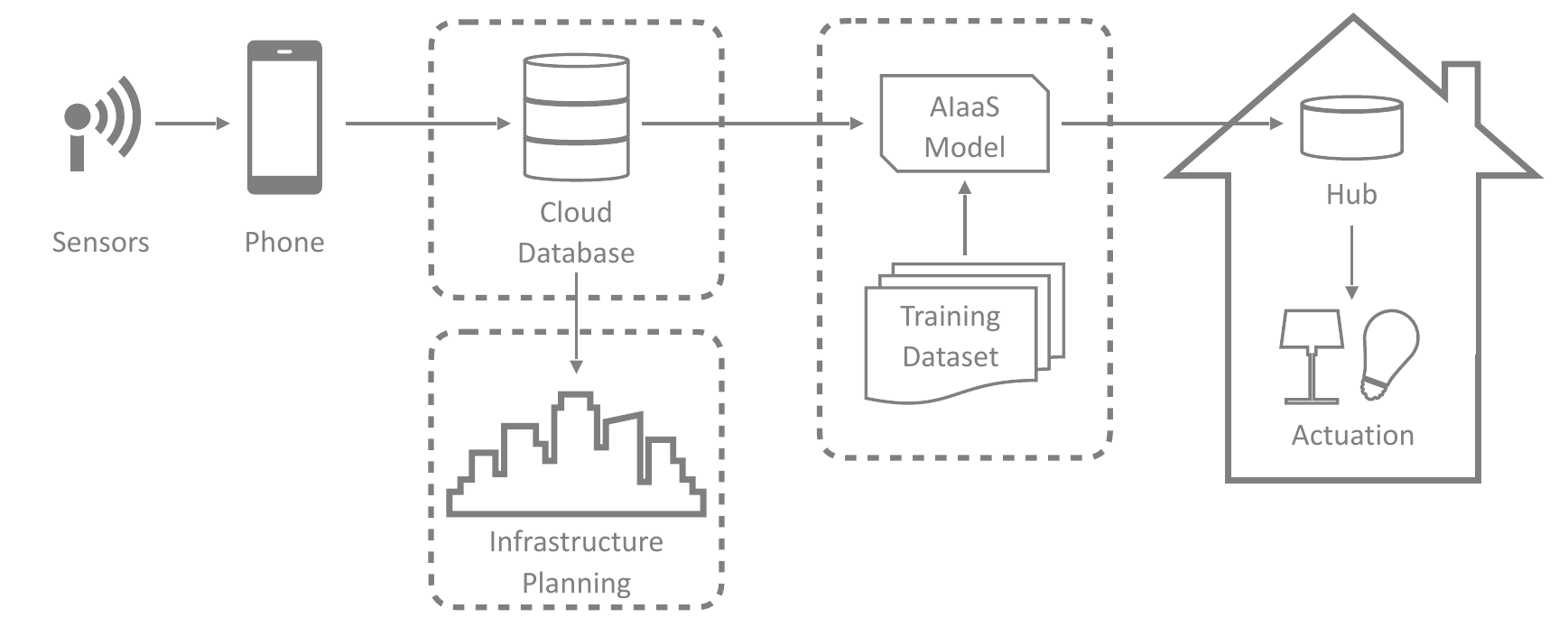}\\
\end{center}
\vspace{-1.5mm}
\footnotesize{
\noindent Figure 1: An example IoT application, where by sensor readings work to trigger an actuation in a smart home, with some data feeding to a city council's planning service. Even a simple scenario involves data flowing across a range of technical and organisational boundaries, as each arrow indicates. Decision provenance works to capture information about these data flows, thereby supporting review.}
}
\end{boxes}
%\end{figure}

Decision provenance is predicated on the idea that it is the flow of data---the interactions between components in a system---that drives system functionality. Its name is in recognition that the actions or decisions taken in a systems context, be they manual (e.g. initiated by users) or automated (e.g. resulting from the application of  an ML model), entails a series of steps that leading up to a particular happening, as well as the cascading consequences. All of these steps entail data flow (at some level) about which information can be recorded. Indeed, this characterisation directly accords with the grand visions of the IoT. 

% \todo{chris can you add examples throughout this para} 
The purpose of decision provenance is to expose \textit{decision pipelines}, by providing records regarding the happenings leading up to a particular decision\slash action, and the cascading consequences. This can be at design time (e.g. by capturing information about a machine learning process, such as the data sources comprising a dataset for model training), run-time (e.g. capturing information about the data flowing to particular entities, and the decisions being made by algorithmic components), or at ``review-time'' (e.g.  capturing information about how systems are audited or investigated). 
In this way, it can operate to provide a broader view of system behaviour and the interactions between the actors involved. 

Decision provenance can assist various reviewability aims in an IoT context. That is, decision pipelines can allow, for example, those building and deploying systems to improve their quality, and to identify points of failure, while empowering others by offering more insight into the nature of the systems and data that is beyond their direct control~\cite{pasquier2017, singh2016b, singh2018}. 

Further, decision provenance, in providing information on data flow, also paves the way for the more proactive accountability measures. This is because the provenance data captured regarding the nature of the data flowing into, within, and beyond a system provides the foundation for particular actions or behaviour to be taken in response. These responses may be human or organisational, or perhaps technical through pre-defined automated, event-based tooling~\cite{singh2016b}. Such mechanisms pave the way for proactive actions to assist compliance and obligation management~\cite{singh2016b}. For instance, information about data flows could be used to trigger particular compliance operations; such as to automatically report data breaches to the relevant authorities; to screen and filter out data based on compliance criteria (e.g. data past an expiry date or use); to not act on inputs that may have come from, or through, an unreliable entity~\cite{singh2015}; or to automatically prevent data flows that are unexpected (in terms of a pre-defined management policy~\cite{singh2016b}). Such approaches may be particularly useful for IoT applications given their potential to exhibit emergent nature.

%Note that provenance will by no means itself solve accountability challenges (nor will any technical measure for that matter).
%However, we present decision provenance to illustrate that how provenance methods show promise in complex systems environments such as the IoT, by exposing and enabling the ongoing monitoring of the relations, interactions and dependencies of such systems where such visibility may not otherwise exist. 

We present decision provenance to illustrate how provenance methods show promise in complex systems environments, such as the IoT.
Provenance will by no means itself \textit{solve} accountability challenges (nor will any technical measure).
However, it does show potential in supporting review, by exposing and enabling the ongoing monitoring of the relations, interactions and dependencies of such systems where such visibility may not otherwise exist. 

% -- including, how and where to store, manage and protect such data; issues of security, trust; how to make such information meaningful to interrogators; to name but a few.  
%such as how to capture and resolve records at different levels of technical abstraction (including that out of band); how capture mechanisms can reliably operate across administrative domains -- including, how and where to store, manage and protect such data; issues of security, trust; how to make such information meaningful to interrogators; to name but a few.  

%\noindent Such records are not only useful for (ex-post) audit and investigation, but can also actively drive or determine interventions and actions.\footnote{In a similar vein to how run-time provenance has been used for detecting and notifying of system faults~\cite{han2017}.

\subsection{Technical challenges}
%Beyond mapping out - is building the reactive response stuff
%Note this is just the tip of the iceberg. Lost of issues of how you collect, trust, manage and store the information. HCI too
%However, it's an important part of supporting reviewability, and thus a rich research area to explore

We have outlined how information on the nature of systems is important for supporting review processes, with provenance representing but one way forward for capturing such information in an IoT context.
Given accountability has received little attention by the technical community, there are a number of research challenges and opportunities for tools and methods that support accountability regimes~\cite{singh2018, singh2019}.

One key consideration is performance and scalability. Record keeping has the potential to be onerous, not least as there may be range of actors interested in reviewing such information. This means that a wealth of technical detail will often be required to satisfy the various aims of the reviews that can potentially be undertaken. It is therefore important to consider carefully what and how much information is likely to actually be needed to provide meaningful accounts of system behaviour {(the `relevance' and `proportionality' dimensions of contextually appropriate information)}. Here, considerations include the overheads imposed on devices, especially given the heterogeneous nature of components and their capabilities in the IoT space. Also crucial for deployment are architectural considerations, including policy regimes that can flexibly determine what, when, where, and how data is captured, stored, and transmitted. Similarly, mechanisms are needed that are capable of working reliably and efficiently across both technical and administrative boundaries, and are able to aggregate or resolve data from across these.

%Another challenge concerns information variety and veracity \changejs{(the `relevance' and  `accuracy' dimensions of contextually appropriate information)}. %hack for now.
Further, for a review to be meaningful, it must be based on complete and reliable information {(the 'accuracy' dimension of contextually appropriate information)}.
This can be a challenge in an IoT context, where components may be operated by different entities, with different aims. 
This means the information relevant for review might come from various sources, be they different technical components (potentially operating at different levels of technical abstraction), by different entities, and potentially requiring supplementation with non-technical (out-of-band) detail. Mechanisms for resolving and aligning captured information is an area requiring consideration. Similarly, the risks and incentives in an accountability context are complex, meaning that the records themselves can pose an organisational burden -- given that information might relate to their responsibilities, and has the potential to be used against them (as part of a recourse mechanism). 
Means for ensuring the integrity and validity of not only the technical information provided, but also the mechanisms that capture, manage and present that information are therefore important considerations.

Related is that capturing information of systems is essentially a form of surveillance. This bears consideration, as the information captured might be inherently sensitive, be it personal data, representing confidential business processes, or generally requiring a stringent management regime. Mechanisms that help ensure that only the appropriate parties are able to view the right information in the right circumstances is an important area for {consideration} and research.

%\notejs{@Chris you might want to rework the next bits re HCI -- these were munged in from things before.}
Usability is another key consideration; as we have made clear for a review to be effective, the information of technical system \textit{must be meaningful} for reviewers  {(the `comprehensible' dimension of contextually appropriate information)}. 
However, one challenge in this space is dealing with the complexity of the technical information (systems provenance information, for example, can quickly become extremely complex~\cite{chen2012, davidson2008, oliveira2017, wang2018}), as well as how detailed technical information can be made accessible and understandable. 
Tools that facilitate the interrogation and representation of technical details will be crucial for facilitating meaningful review in an IoT context, though further work is required to explore the presentation of technical details for accountability purposes.

As such, there are real opportunities for human computer interaction (HCI) research methods to be employed and extended to assist with the usability and interpretability of audit information. Through such means, the tools and techniques available can better support the aims and literacies of the various actors (technical experts, regulators, users, etc.) in conducting a review. Though there has been some consideration of this in certain contexts (e.g. in some specific provenance cases~\cite{chen2012, li2012, oliveira2017,bachour2015,schreiber2017,norval2019}), there appears real scope for more general explorations into such concerns. The development of standards and ontologies could assist (e.g. \cite{ujcich2018} describes an ontology for the GDPR) with both the management and interpretation of information for review.

%!TEX root = main.tex
\section{Reviewability in practice: a smart city}

% \begin{itemize}
%     \item Greater knowledge of the sources and veracity of data leading to a given decision or action
%     \item Identify changes and conflicts within the IoT deployment at run-time
%     \item Assisting with investigations to identify initial points of interest
% \end{itemize}

%\todo{4. This might be slightly out of scope for the chapter (especially considering how section 5 has focused mainly on technical measures), but since the authors speak of both technical and organisational workings of IoT systems, it might be worth at least a brief discussion on how organisational arrangements can also make a difference. One such possibility is – me being a lawyer – defining the reviewability duties of different parties in a contract. This could be covered by, for example, mentioning in the case study in section 6 that the Decision Provenance record was in place because the vehicle manufacturer has secured such an obligation throughout the supply chain in the first place. This is of course just a suggestion and it’s up to the authors how this should be discussed, but there should be at least some indication about possible organisational approaches.}

We now use an example of a smart city, a common IoT scenario, to illustrate at a high level how methods supporting %socio-technical 
reviewability relate to broader accountability concerns. 

Smart cities aim to make ``better use of the public resources, increasing the quality of the services offered to the citizens"~\cite{zanella2014}, and the IoT is often seen as integral to achieving such functionality~\cite{liu2019}. Some commonly considered examples include the automation of home security~\cite{dlodlo2016}, traffic rerouting~\cite{misbahuddin2015, saha2017, xiong2012}, vehicle control~\cite{mckee2017, xiong2012}, and disaster management~\cite{asimakopoulou2011, boukerche2018, dlodlo2016}.
%\notejs{these are correct, but super boring. Can you get some more modern examples, including that with more automation... even these examples, rerouting traffic on demand. Also I think in this section we should have scenarios that involve users and some dynamism -- much of these examples can be hardwired and case specific. We need specific users that hit harder on the complexity/unknowns etc}.
Such deployments may involve data and interactions from a wide range of sensors (infrastructural and/or mobile), devices and actuators, online services (cloud storage, AIaaS), etc, which may personally, privately, or publicly operated.
Naturally, large number of actors may be involved, exchanging and processing data in order to achieve some overarching functionality.
%Such deployments represent a wide range of system components and actors, may include the use of cameras, microphones, and GPS, and involve device manufacturers (sensors, vehicles), online services (cloud storage, MLaaS), and local authorities (city planning, department for transport) and of course, citizens. \notejs{split this out -- you have a wide range of devices - in infrastructure, deployed, and mobile; then you have different actors}

Importantly, while the boundaries of an IoT application may appear clear and well-defined (at least to those that operate them), in practice they will often interact (communicate and exchange data) with other systems -- either to provide the expected functionality, or perhaps to be used and reused in ways beyond what may have been originally intended.
For example, information may be collected independently by home, vehicle, and mobile sensors, which might interact with deployments within public and private spaces (e.g. for automation, security, insurance, billing), which may in turn feed data into broader algorithmic systems and services (energy grid, emergency services, public transport, etc), all of which may result in decisions and actions with real-world consequences.

In these complex systems arrangements, any number of things could potentially go wrong; 
a data breach could see the release of personal data, automated actions could be erroneously taken, inferences could be made on incorrect information, and so on.
Yet, given the ability for such issues to cascade throughout an IoT ecosystem, the true extent of the impact may not always be clear.
As we have argued, the ability to review such systems (and their role at each step in the process) may be instrumental in identifying who was involved, how such incidents occurred, and any `knock-on' effects.
We now explore how such an investigation might take place, and how the ability to review IoT ecosystems can be leveraged to provide legal and operational benefits alike.

\subsection{A traffic incident}

%\notejc{one thing about all of this is that it's almost entirely about technical logs, with the exception of information about the training dataset for the vision system. We really could do with something involving records about organisational processes. Maybe records show that the ambulance deployment model was supposed to be checked every week or two by its developers to make sure it was all fine and providing them with the patterns of deployment that they'd expect but records show that the check hadn't been carried out?}

Consider a traffic incident occurring in a smart city. 
While walking to a sporting event, a pedestrian crosses the road at a busy set of traffic lights and is struck and injured by a vehicle. 
A witness immediately calls an ambulance, which is dispatched to the scene of the incident, but takes a very long time to arrive. 
Given the nature of the smart city ecosystem, {a regulator commissions an investigation} %investigation takes place 
to discover what parts of the IoT infrastructure {may have} contributed to the incident, what went wrong, and who (if anyone) may be liable.

Speaking to those involved, the investigator gains three initial lines of inquiry:
(i) the injured pedestrian claims that the driver ran a red light; 
(ii) one witness commented on how dark the streetlights were, despite it being a busy area with lots of vehicles and pedestrians, and questioned whether that might have affected visibility; 
(iii) the ambulance driver noted that they had been in that area earlier, but had been redirected away shortly before the incident occurred. 
In all cases, these three potential threads entail interactions with the smart city's infrastructure.

\subsubsection*{What role did the driver play?}
The injured pedestrian had indicated that the driver ran a red light.
However, the driver claims that they were not at fault, the argument being that their vehicle is one by CarNet---a highly popular manufacturer of networked `smart cars'---and could therefore not have breached the traffic rules. 
CarNet is known to use complex AI systems to detect and inform of hazards on the road, in collaboration with a mesh network of all nearby CarNet vehicles, and autonomously applies the brakes and other countermeasures to avoid accidents. 
Since the vehicle did not detect the hazard (the pedestrian), the driver believed that they couldn't have been at fault, and that the pedestrian must have suddenly and unexpectedly attempted to cross the road at a dangerous time.

Given that CarNet have built their systems to be reviewable, they retain extensive records about the design of their cars' systems, as well as ensuring each car produces logs of their operation. 
Log files containing certain information on how systems are functioning are sent to CarNet (including whether systems are operating correctly and sensors are reporting in line with expected parameters), however, some data (such as personal data, including information on vehicle movements and camera feeds) %---such as information on vehicle movements---
remains locally stored on each car. 
{CarNet has also secured contractual obligations throughout their decision supply chain, ensuring that third-party suppliers (on which their vehicles depend) also have record keeping practices in place. %record keeping practices are also in place by the third-parties suppliers on which their vehicles depend. 
The result is a high degree of oversight and reviewability over how CarNet's vehicles operate on the road.}

The investigator reviews CarNet's records, which indicate that the car's systems and sensors were fully operational, and logs retrieved from the car in question indicate that there were four CarNet vehicles nearby at the time of the incident, which shared information about potential hazards at the particular area where the incident occurred.
Looking more in-depth at the operational logs retrieved from the car (including camera footage fed into the car's AI system, and network communications sent from other vehicles), the investigator determines that the traffic light was indeed red -- however, none of the vehicles detected any pedestrians or other hazards near to, or present on, the road.

%\notejs{should CarNet self investigate? or should this be the investigator. this shows CarNet in a bad light, could imagine them seeking to withhold info}\notejc{there's also something to consider here about where the logs are recorded - if logs about driving are recorded by CarNet then that is a pretty big privacy/data protection issue and we presumably don't want to be framing reviewability along those lines. It should be possible to do reviewability in as privacy-preserving a way as possible. Should we say that the investigator gets logs from the car, rather than CarNet? I've made a suggestion above}\notecn{I've implemented these changes, as well as reframing tweaks suggested by Jat.}

Through reviewing the data from the car, the investigator also learns that CarNet vehicles work by using an object recognition model which takes camera footage from their vehicles and determines the location of pedestrians and other vehicles on the road.
This model is built and operated by CloudVision through their cloud-based AIaaS service, though the vehicle uses a locally stored version which is periodically updated from CloudVision's servers.
{The investigator notes that the model has not been updated in seven months -- and this was initiated by the mechanic, the last time the car was serviced.} 
%\notejs{there's an opportuntiy to say, e.g., the last 10 updates failed, in which the lighting problem was fixed.. that way cloudvision can say well if you had the update you'd have been fine, butfor the lowlight situation..  wdyt?}\notecn{That works. Could also say that updates had been released but the model didn't automatically update. CarNet's processes relied on the user being active on that front.} 
Decision Provenance records from the car demonstrate that the object recognition model had failed to detect either the red light or the pedestrian from the camera feeds sent to it (as indicated by records of the model's outputs). 
Speaking to CarNet, while they concede that the records show that their system had failed to detect the hazard prior to the incident, {they argue that the driver should have been regularly updating their AI software in order to improve how it operates.}
They also argue that their interventions are meant as a last resort, that their systems are merely to support drivers and that the driver should have been paying more attention and taken action to avoid the incident.

%\notejc{I like this - very Tesla-esque denial of responsibility. Almost makes me want to rename CarNet to Edison Cars or something}\notecn{Ha. Tesla was partially the motivation behind this}

%\notejs{it's a bit unrealistic to rely on cloud connectivitiy for real-time AI}\notejc{we can say that the system checks for an updated model each morning and uploads DecProv data when returned to base, or for investigation?} \notecn{implemented}

%\notejs{Do they do this, or an investigator? regulator? @Jen - views?}\notejc{I would imagine that it would be a regulator - actually I think most of the investigation would be carried out by a regulator}\notecn{I've tried to place the `investigator' in the lead for most of these reviews. Happy to rename `regulator' if it helps things.}

The investigation then turns to CloudVision's role in the incident.
{Given their contractual obligations with CarNet to implement reviewability, CloudVision retains extensive documentation and logs over their service. This allows the investigator to perform}
%The investigator performs 
an in-depth review of their object recognition model through using techniques discussed in the algorithmic accountability literature (e.g.~\cite{diakopoulos2016}), including looking at the `datasheets'~\cite{gebru2018} of the datasets used to train the model{, and the `model card'~\cite{mitchell2019} for its performance benchmarks across different conditions.}. %\todo{@chris - add model cards}\notecn{done}.
The investigator finds that the model was not adequately trained to handle low light scenarios, {and its ability to accurately recognise objects decreased significantly when presented dark images} -- indicating that
% indicates
their models would not be properly equipped for the low levels of street lighting that were observed at the time of the incident.

CloudVision were contacted, but claim that they were not responsible; they have their own conditions of use which specify that their models should be updated on a regular basis.
CarNet's vehicles should have been frequently updating to the latest version automatically, as oppossed to relying on the driver manually looking for updates. %by having their own conditions of use stating {that their model should be updated on a regular basis, and that CarNet's vehicles should be automatically updating to the latest version and not relying on the driver manually looking for updates. 
Their change logs {and model cards} indicate that a recent update improved the model's ability to classify objects in low light conditions, but their guidance for developers still recommend that other complementary or redundancy measures be employed where situations of low light may occur.
%their services should not be used for high-risk scenarios in low light conditions, and in their guidance for developers, recommend that other complementary or redundancy measures be employed where situations of low light may occur. 
CloudVision also produce records showing that these usage restrictions were reflected in the contract they have with CarNet, and that they previously discussed issues of low light with CarNet, but the distributed nature of the model (i.e. they run locally, on-vehicle) meant that CloudVision had limited oversight over when and how it was used once deployed.

%\notejs{This needs explanation -- it's not immediately obvious why this would be an improper usage by the vehicle provider -- surely usage at such a scale would have some procurement contract, meaning CloudVision would at least be aware that CarNet was using their app} \notecn{I thought this could be an indication that non-technical processes had broken down (i.e. a service being used in a way that was unintended), or maybe they were just trying to lean on legal terms to absolve themselves, but this might not really work in practice as you say. Happy to remove reference to them saying it shouldn't have been used if that makes things a bit more robust?}\notejs{It's an interesting point. I like saying things are misuse. But it'll need to make clear how they didn't use it as intended? I mean the usage of the service would almost certainly mean that CloudVision knows of CarNet as a customer... and cars need to drive at night... so... where's the misuse}\notejc{could say that CloudVision advise against using the system at night but CarNet perhaps haven't made that clear to their customers} \notecn{I've tweaked it a bit to try and reflect this. Does it work better? The fact that we now say that the model is locally deployed does introduce the potential that CloudVision had limited oversight over how and when their models were used (while giving them a good excuse to try to wash their hands of it)}

Having holistically reviewed the records and logs made by CarNet and CloudVision about their processes and systems, the investigator concludes that while the driver may have been at fault, there was also a failure in CarNet's system which did not employ automated braking or countermeasures since the pedestrian and the red light were not detected.
This was determined to be due to CarNet vehicles being able to function using an {an outdated version of} CloudVision's object recognition service which did not operate appropriately in low light conditions{, and identified failures in CarNet's processes which relied on drivers manually updating their vehicle's software}. However, the abnormally low levels of street lighting on the road at the time of the incident may have also contributed.

\subsubsection*{Why were the street lights dimmed?}
The investigator learns that the management of the street lighting on the road of the incident was recently outsourced to SmartLight Inc., a supplier of automated street lighting solutions which vary the brightness of the lights according to levels of activity (busyness) in the area. 
On reviewing the street lighting system, the investigator discovers that the service works by retrieving external information about the number of pedestrians and vehicles present on the street, which is obtained through CloudMap's popular mapping service.
Through SmartLight's log files, the investigator determines that the mapping service had indicated that no vehicles or pedestrians were present at the time of the incident, which led to SmartLight's system determining that the street lights should be dimmed in line with the Council's desire to conserve electricity.
However, the camera footage from the vehicle clearly shows that the area was busy, and therefore CloudMap's information was incorrect.
Asking SmartLight about the reliability of CloudMap, they explain that they have never had any past problems with the service, and that they had therefore not implemented any back up data sources or contingency plans for when the mapping service fails to supply accurate information.

In considering  CloudMap's mapping service, the investigator can see that their pedestrian and vehicle density information is `crowdsourced' through the mobile devices and vehicles that have the mapping application installed. This data is aggregated to determine where congestion is high and where delays are likely.
CloudMap is approached regarding the discrepancy, who (after checking their release schedule) inform the investigator that they had temporarily rolled out an update which appears to have negatively affected their cloud databases and mapping tool.
This resulted in incorrect information relating the congestion of certain roads, including the one on which the incident occurred.
{On probing, CloudMap reveal that their access logs show that} this incorrect information was also retrieved by EmergenSolutions Ltd.---an organisation known to be responsible for coordinating and distributing emergency service vehicles across the city---and that they may have also been affected by the issue. 
%Looking at the CloudMap access logs for this street, they note that the incorrect information was also retrieved by EmergenSolutions Ltd.---an organisation responsible for coordinating and distributing emergency service vehicles across the city---and that they may have also been affected by the issue. 
%\notejc{looking at access logs to see who else was using it in that street seems like too much of a privacy violation to me. Maybe we can just say in the next bit instead that EmerSolutions also used the same mapping service?}\notecn{Does this fix it? I.e. CloudMap disclosed this to the investigator, that the investigator may not have had access themselves. This seems like something that might come out of a regulatory investigation?}
%\notejs{again, does this need tweaking? Perhaps better is to find out ther was no traffic, then checking cloudmap it becomes clear an update was isseued (or even better - the update was 'detected' from looking at a system relying upon it}
%EmergenSolutions had previously come to the attention of the investigator through their part in the emergency service's response to the incident. \notejs{this para shows the issues nicely }

\subsubsection*{Why was the ambulance delayed?}

The investigator turns to how this information was used by EmerSolutions. 
Records help reveal that the area where the traffic incident occurred had been classified as `low risk' for emergency services. This then led to an automated decision to redirect ambulances away from the area shortly before the incident took place.
This `low risk' classification resulted from three data sources;
(i) information from CityMap reporting that there was low congestion in the area;
(ii) data provided by CarNet indicating that few pedestrians or vehicles were present on the street, and
(iii) historic information about emergency service call-outs showing that the street was not a `hot spot' for incidents.

On contacting EmerSolutions, the organisation emphasises that they have redundancy measures in place, and that their planning system distributes emergency response vehicles based on these disparate sources of information.
Probing further, the investigator discovers that EmerSolutions had a process in place to adjust the impact of historic information when large-scale events are on, such as sporting and music events -- however, this process had not been followed for several weeks, and no record had been made of the major sporting event that the pedestrian was walking to. This was not reflected in their models, and thereby hindered the speed of response by the emergency services. Further review of the system indicates that had this process been followed, the diversion of emergency service vehicles from the area would not have occurred and an ambulance would have been closer to the incident when it occurred, as would have been appropriate and expected when a major event takes place.

%Given what the investigation has uncovered about the incident---that (i) CarNet's ability to detect pedestrians was affected by the low street lighting; (ii) CityMap's services had been temporarily affected by the update; and (iii) operational processes were not followed by EmerSolutions---the investigator now feels that their review of the IoT ecosystem has identified multiple failures that contributed to the incident.

\subsubsection*{Putting together what happened}
%\notejs{It does feel a bit flat -- like it's a complex scenario, but reviewability doesn't really bring about a happy ending (the implication is someone died!!} 

%\notejs{Can this be strengthened -- i.e. what would be the outcome if there WASN'T reviewability. I think we need to hit home on something strong} \notecn{I've tried to tweak the wording throughout so that there are multiple instances where these insights only came out through the reviewability measures previously put in place by the entities. Is it worth saying in this section that without reviewability, this investigation may not have identified the underlying points where failures occurred and how they could be addressed?}

This example indicates that even incidents that appear simple can be complex, and that deployments designed to facilitate review will not only be important, but necessary for supporting investigation. The ability to review the technical and organisational workings of complex IoT ecosystems can provide both legal and operational benefits to the various entities involved. 

From a legal standpoint, effective record keeping and logging will assist investigators in mapping out what happened and where issues originated. %\notejs{better would be to show *how* they assisted, referring back in only a few words how this relates}. 
For instance, the scenario shows that reviewability allowed the investigator to determine that (i) CarNet's ability to detect pedestrians was affected by the low street lighting {and questionable software update procedures}; (ii) CityMap's services had been temporarily affected by an update; and (iii) EmerSolutions failed to follow their own operational procedures. As such, because systems and processes were designed to be reviewable, the investigator was able to identify multiple issues, with several entities contributing to the incident throughout the IoT ecosystem -- reflecting that there may be several points of failure in any ecosystem, and further that apportioning blame will often be complex in reality. That said, mechanisms that shine light on the happenings of systems means that the appropriate legal consequences, in terms of assigning liability and pursuing redress, are able to follow from the information obtained through review.

%\notejs{the concept re this next para is good, but needs to say explicitly how each actor will impove their processes}
Operationally, reviewability has also allowed the organisations involved in this example to obtain insights into weaknesses or potential misuses of their systems.
As such, they are able to improve their operational processes and prevent problems from re-occurring in future.
Such information may also feed into their processes for ensuring and demonstrating legal compliance, by showing that they are quick to identify and act upon failures. %\notejs{I think unpack this -- gotta emphasise to stop issues again}.
For example, based on the investigations that took place, {CarNet may choose to change their processes around software updates to have their vehicles automatically look for updates every night.}
%CloudVision may decide to adjust how their image recognition system performs under low lighting conditions, introducing new datasets to ensure that the system can accurately classify darker images. %%% Note: I took this out as it seems less of an issue on their part given that they fixed it. But we also say more below about how they might want to better control how their clients use their models.
CloudMap may re-assess their processes surrounding their deployment of updates -- perhaps better staging releases or having a more rigorous testing process to prevent such issues {from being deployed in production}. Both CloudVision and CloudMap may also choose to more closely monitor what kinds of applications and clients are using their services, and work with those clients in more high-stakes scenarios to implement bespoke services or processes which ensures that a high level of functionality is maintained in critical situations. Similarly, SmartLight may review their role in the incident, and set up contingency plans for when the mapping service fails -- perhaps using a second mapping service, or other sources of information (such as historic data from past usage logs), to ensure that the lights aren't dimmed on streets that are known to likely be busy.
EmergenSolutions may learn from the incident that relying too heavily on CityMap's services may have implications for system resilience, and that their failure to follow internal procedures can lead to issues with potentially significant consequences.

In the cases above, reviewability plays an important role for identifying and addressing issues in socio-technical systems. Without access to the comprehensive records and logs needed to holistically review the operation and interaction of the various component parts of IoT ecosystems, it would be extremely challenging to reliably identify where problems arise and where failures occur. Without oversight of the various entities involved (as well as their roles in the incident and their organisational processes), it would difficult, perhaps impossible, to act on the investigation and take next steps. As we have shown, implementing reviewability---through comprehensive technical and organisational record keeping and logging mechanisms---offers significant benefits in terms of the improving of accountability in these complex IoT ecosystems.

\section{Concluding remarks}
%\notejs{We needed a conclusion.. this is more a placeholder.. [old text in comments. Feel free to edit at will]}
%\todo{Please review}

There is a push for stronger governance regimes, as the IoT continues to become increasingly interconnected, algorithmic, and consequential.
However, the complexity of the socio-technical, systems-of-systems that comprise the IoT  introduces opacity, which poses real challenges for accountability.

We have argued that a key way forward is approaching accountability through the concept of reviewability. This involves mechanisms, such as logging and record-keeping, that provide the necessary information to allow technical systems and organisational processes to be comprehensively interrogated, assessed, and audited. The aim is to provide for a targeted form of transparency, that paves the way for meaningful accountability regimes. 

There appears real potential for technical methods, such as those provenance-based, to capture information of the systems that form the IoT. Such information can work to facilitate processes of review, however, more {research} is required. 
{Though here we have focused on some technical capture mechanisms, the wider concerns are often socio-technical. Reviewability will therefore require a broad range of technical and non-technical record keeping measures alike -- from the commission and design of the IoT infrastructure, all the way to its continued operation. In other words, {implementing effective} reviewability requires a holistic approach, encompassing a range of perspectives that expose the contextually appropriate information necessary to assess the wider contexts in which the system is deployed.}
%\cut{Tackling the challenges of reviewability is beyond that just technical, and requires a holistic approach, encompassing a range of perspectives.}
There are legal, compliance and oversight considerations, issues around incentives and organisational processes, the usability and relevance of the information driving review, as well as technical aspects concerning system design, development and monitoring, to name but a few. Realising greater accountability involves considering the relationships and interplays between these concerns, and more. 

%\notecn{I wonder if it makes sense to have a paragraph in here about how, while this stuff needs work, there are advantages to start recording processes \emph{now} (I think we had a sentence as such saying that in the DecProv paper). Means that there's a more tangible take-away for readers who are developers and interested in improving their processes, and rounds off the paper nicely for any devs reading. I can have a go when I get to this point in the pass through.}
In discussing IoT reviewability, this chapter seeks not only to draw attention to issues regarding the complex, interconnected, data-driven and socio-technical nature of IoT systems, but also aims to encourage more attention and collaborative actions towards improving IoT accountability regimes. These concerns will become all the more pressing as the IoT continues to pervade our world.
And despite the challenges in bringing about reviewability `at scale', there are real and immediate benefits in employing any mechanisms that support review \textit{now}. 
Even %technical and organisational - looking at it, we've said it plenty of times.. easier to read now without it
mechanisms that only support internal review (i.e. within an organisation or development project) can help in the management of systems, processes, and obligations, in addition to assisting compliance and providing evidence demonstrating good practice. In short, any reviewability undertakings can work to pave the way towards an IoT that is more understandable, transparent, legally compliant, and therefore accountable.

\section{Acknowledgements}

We acknowledge the financial support of the Engineering \& Physical Sciences Research Council (EP/P024394/1, EP/R033501/1), University of Cambridge, and Microsoft via the Microsoft Cloud Computing Research Centre. 

%\todo{Check the refs - some are messed up}

\bibliographystyle{vancouver-modified}
\bibliography{references.bib}

\end{document}